\documentclass[aps,amsmath,amssymb,superscriptaddress,showpacs,twocolumn,floats,epsf]{revtex4}
\usepackage{amsfonts}
\usepackage{graphics}
\usepackage{graphicx}
\usepackage{hyperref}
\usepackage{color}

\begin{document}

\title{Chiral Symmetry Breaking and the Quantum Hall Effect in Monolayer Graphene}

\author{Bitan Roy}
\affiliation{Condensed Matter Theory Center and Joint Quantum Institute, University of Maryland, College Park, Maryland 20742-4111, USA}

\author{Malcolm P. Kennett}

\affiliation{Department of Physics, Simon Fraser University, Burnaby, British Columbia, Canada V5A 1S6}

\author{S. Das Sarma}
\affiliation{Condensed Matter Theory Center and Joint Quantum Institute, University of Maryland, College Park, Maryland 20742-4111, USA}


\begin{abstract}
Monolayer graphene in a strong magnetic field exhibits quantum Hall states at filling fractions $\nu = 0$ and $\nu = \pm 1$ that are not explained within a picture of noninteracting electrons. We propose that these states arise from interaction-induced chiral symmetry-breaking orders. We argue that when the chemical potential is at the Dirac point, weak on-site repulsion supports an easy-plane antiferromagnet state, which simultaneously gives rise to ferromagnetism oriented parallel to the magnetic field direction, whereas for $|\nu|=1$ easy-axis antiferromagnet and charge-density-wave orders coexist. We perform self-consistent calculations of the magnetic field dependence of the activation gap for the $\nu = 0$ and $|\nu| = 1$ states and obtain excellent agreement with recent experimental results. Implications of our study for fractional Hall states in monolayer graphene are highlighted.  
\end{abstract}


\maketitle

There have been theoretical proposals that strong electron-electron interactions can lead to a variety of broken-symmetry phases in monolayer graphene (MLG) \cite{herbut-juricic-roy, dima, drut-lahde}. However, their experimental detection has remained elusive, with graphene seemingly behaving as a weak-coupling system down to the lowest achievable densities \cite{sankar-Coulomb}. However, the formation of interaction-driven ordered phases can be catalyzed by quantizing magnetic fields 
that enhance the effect of electron-electron interactions by developing a 
set of highly degenerate Landau levels (LLs) \cite{gys-mir-gor-shov, KYang-review, goerbig-review}. At low magnetic fields 
Hall plateaux are observed for $\nu=\pm (4 n+2)$, which can be understood in a non-interacting electron picture as 
arising from \emph{fourfold} valley and spin degenerate two dimensional Dirac fermions \cite{qhe-graphene-1,qhe-graphene-2,sharapov}. 
At higher magnetic fields, additional plateaux appear at $\nu=0, \pm 1$ and $\pm 4$ \cite{Kim-LLsplitting}. 
The $\nu=\pm 4$ plateaux most likely arise from single-particle Zeeman splitting of the LLs, whereas the appearance of $\nu=0, \pm 1$ states strongly suggests that the fourfold valley and spin degeneracy is lifted by interaction driven broken symmetry phases within the zeroth LL (ZLL) \cite{Kim-LLsplitting, novoselov-PNAS, yacoby-PRB, kim-recentscaling, EAndrei-1, FQHE-Kim-NatPhys, EAndrei-FQHE, FQHE-Yacoby-Science}.  

The ZLL is distinct from other LLs in MLG because it is simultaneously valley and sublattice polarized.  Proposed broken symmetry phases
that can cause splitting of the ZLL \cite{dassarma-yang-macdonald, moessner, fuchs, nomura-macdonald, herbut-qhe, herbut-so3, jung-macdonald, semenoff-zhou, gusynin-miransky-PRB, barlas-review, kharitonov} fall into two classes: $(i)$ chiral (sublattice) symmetry breaking (CSB) orders, such as antiferromagnetic (AFM) and charge-density-wave (CDW) order \cite{dima, herbut-juricic-roy, herbut-qhe, herbut-so3}, and $(ii)$ a ``valley-odd" quantum Hall ferromagnet (QHFM) \cite{semenoff-zhou, KYang-review, barlas-review}. In this Rapid Communication we point out that it is important to consider the influence of \emph{all} of the filled LLs, not just the ZLL, in order to determine which specific symmetry-broken ordered phase is favored at low temperatures. For a magnetic field of strength $B$, the filled non-interacting LLs at $-\sqrt{2 n B}$, with $n=1,2, \cdots$, are \emph{pushed down} to $-\sqrt{2 n B+\Delta^2_c}$ in the presence of a CSB order ($\Delta_c$) \cite{catalysis-original, klimenko}, while for QHFM order ($\Delta_Q$) they split into $-\sqrt{2 n B} \pm \Delta_Q$, which therefore lowers the energy only by splitting the ZLL. Thus, the total energy of the filled sea of Dirac LLs is maximally lowered by the formation of CSB orders. This is strikingly different from the situation in non-relativistic two-dimensional systems, such as GaAs, where Hall states appear only within the first few LLs, so filled LLs play a minor role at high magnetic fields where the LL coupling is negligible \cite{girvin}. Hence, for experimentally accessible fields ($B \lesssim 50$ T), when $\sim 100$ filled LLs lie inside the ultraviolet cutoff $(\Lambda)$ for the effective Dirac dispersion, it is not sufficient to make a ZLL approximation for MLG to distinguish different forms of ordering.  Even though CSB ordering dominates in the ZLL, QHFM ordering plays a role in removing the \emph{valley degeneracy} from higher LLs $(n \geq 1)$ \cite{barlas-review, sheng}, e.g. yielding Hall plateaux at $\nu=\pm 3$ \cite{nu3-firstexp}.

The dependence of the quantum Hall activation gaps on magnetic field at fillings $\nu =0,\pm 1$  has been established via 
multiple experimental techniques: compressibility \cite{yacoby-PRB}, transport \cite{kim-recentscaling} and capacitance \cite{novoselov-PNAS}. \emph{Together} these experiments show that in perpendicular magnetic fields the activation gap for the $\nu=0$ Hall state exhibits a crossover from linear \cite{kim-recentscaling} to sub-linear \cite{ novoselov-PNAS} to an almost $\sqrt{B}$ \cite{yacoby-PRB} scaling (see Fig. 1). A simple estimate of $\sqrt{B}$ scaling due to the long-range Coulomb interaction does not capture this behavior. These results motivate our exploration of the nature of the underlying broken symmetry phases within the ZLL of MLG in a magnetic field, in which we take into account the \emph{finite range} components of the Coulomb interaction. 

Our main results are as follows: \\ 
$\bullet$ We explain the quantum Hall states at $\nu = 0$, $\pm 1$ as arising from interaction driven CSB orders induced by a magnetic field. \\
$\bullet$ We calculate the magnetic field dependence of the activation gap self-consistently and obtain quantitative agreement with 
experiments at $\nu = 0$, $\pm 1$ and show that conflicting scalings of the gap with magnetic field observed in different experiments \cite{kim-recentscaling,novoselov-PNAS,yacoby-PRB} can be unified within a single framework. \\
$\bullet$ Our proposed scenario is consistent with experiments performed in tilted magnetic fields \cite{kim-recentscaling}.

The strongest short-ranged component of the Coulomb interaction in MLG is most likely the onsite repulsion \cite{katsnelson} which has
been predicted to result in an AFM state for fermions on a hexagonal lattice at strong coupling \cite{herbut-juricic-roy, herbut-assaad}. 
In pristine graphene with $B=0$, the interaction strength is insufficient to induce this ordering, but at non-zero field, AFM ordering can take place even for infinitesimal onsite interactions and supports a $\nu=0$ Hall state \cite{herbut-qhe, herbut-so3}. However, the Zeeman coupling ($\lambda$) of the electron spin to the field projects the AFM order ($\vec{N}$) onto the \emph{easy-plane} perpendicular to the applied field and simultaneously supports ferromagnetic (FM) order ($m$) parallel to the field \cite{herbut-so3, roy-BLG}.

We now describe our theory. In the presence of AFM and FM orders the Dirac LLs have energies $\pm E_{n,\sigma}$, where \cite{herbut-so3}
\begin{equation}\label{DiracLL}
E_{n,\sigma}=\sqrt{N^2_\perp+[(N^2_3 + 2 n B)^{1/2} + \sigma (m+\lambda)]^2}.  
\end{equation}
$N_3 (N_\perp)$ is the easy-axis(-plane) component of the N\'{e}el order parameter respectively. $\sigma=\pm$ represents the 
two spin projections. The areal degeneracy of the LL is $D = (2-\delta_{n,0})/(2 \pi l^2_B)$ for $n=0, 1, 2, \cdots$ \cite{supplementary}. At half-filling all the LLs at negative (positive) energies are completely filled (empty). Allowing for both AFM and FM orders, the free energy is \cite{herbut-so3, roy-BLG}
\begin{eqnarray}
F_0=\frac{N^2_\perp+N^2_3}{4 g_a}+\frac{m^2}{4 g_f}-D \sum_{\sigma=\pm} \left[ \frac{1}{2}E_{0,\sigma} + \sum_{n \geq 1} E_{n,\sigma} \right],
\end{eqnarray} 
where $g_a$ ($g_f$) are the short-range components of the Coulomb interaction, such as on-site Hubbard repulsion, that support AFM (FM) order \cite{comment-1, supplementary}. The sum over $n\geq 1$ in $F_0$ is the contribution from \emph{filled} LLs. For any non-trivial Zeeman coupling (i.e. $\lambda \neq 0$) $F_0$ is minimized when $N_3 \equiv 0$ \cite{herbut-so3, roy-BLG}. Therefore, the Zeeman coupling restricts the AFM order to the easy-plane and simultaneously allows FM order parallel to the magnetic field. Taking $N_3=0$, and minimizing $F_0$ with respect to $N_\perp$ and $m$ leads to the coupled gap equations 
\begin{eqnarray}
\frac{1}{g_a} & = & \frac{B}{\pi} \sum_{\sigma=\pm} \left[\frac{1}{2E_{0,\sigma}} +  \sum_{n \geq 1} \frac{1}{E_{n,\sigma} } \right], 
 \label{gapeq1a}  \\
\frac{1}{g_f} & = & \frac{B}{\pi} \sum_{\sigma=\pm}\left[ \frac{(m+\lambda)}{2E_{0,\sigma}} + \sum_{n \geq 1} 
\frac{(m+\lambda) + \sigma \sqrt{2 n B}}{E_{n,\sigma}} \right]. \nonumber \\
& &
\label{gapeq1b}
\end{eqnarray}
FM order splits all the filled LLs, including the zeroth one, while easy plane AFM order pushes down all the filled LLs as well as splitting the ZLL. Thus the contribution from the filled LLs with $n \geq 1$ in the first (second) gap equation add up (cancel). Consequently, Eq. (\ref{gapeq1b}) is free of divergences, but Eq. (\ref{gapeq1a}) exhibits an ultraviolet divergence which can be regularized by introducing $\delta_a=\pi \big[(g_a \Lambda)^{-1}$ $-(g^a_c \Lambda)^{-1}\big]$ \cite{herbut-qhe, roy-scaling}, where $(g^a_c)^{-1}=\int^\infty_{\Lambda^{-1}} ds/s^{3/2}$ is the zero magnetic field critical onsite interaction strength for AFM ordering \cite{herbut-juricic-roy, herbut-assaad}. The inclusion of filled LLs ($\sim 100$) within the cutoff ($v_F \Lambda \sim 3$ eV) is sufficient to capture strong LL mixing  in graphene, but the physical observables ($N_\perp, m$) remain independent of $\Lambda$, so that a continuum description is meaningful.

\begin{figure}[htb]
\includegraphics*[width=8.5cm]{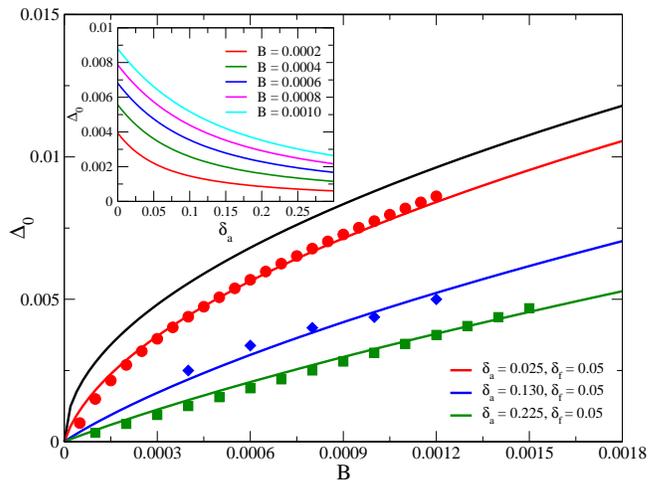}
\caption{(Color online) Scaling of pure AFM order (black) at zero magnetic field criticality ($\delta_a=0$). Red, blue, and green points show the observed gaps in Refs. \cite{yacoby-PRB}, \cite{novoselov-PNAS}, and \cite{kim-recentscaling} respectively and the lines show fits to the $\Delta_0$ data. The magnetic field ($B$) is measured in units of $B_0 \sim \Lambda^{2} \sim 10^4$T, the field associated with the lattice spacing. $\Delta_0$ is measured in units of $E_c=v_F \Lambda$. Inset: Dependence of $\Delta_0$ on $\delta_a$ at several different values of $B$ and for $\delta_f(=\pi/g_f)=0.05$.}
\label{fig:Fits}
\end{figure}

With underlying easy-plane AFM order the total gap for the $\nu=0$ Hall state is $\Delta_0=\sqrt{N^2_\perp+(\lambda+m)^2}$. 
After applying the regularization \cite{supplementary} we solved Eqs.~(\ref{gapeq1a}) and (\ref{gapeq1b}) numerically to 
find $N_\perp$, $m$, and $\Delta_0$, and fitted our results to data from Refs.~\cite{yacoby-PRB, novoselov-PNAS, kim-recentscaling}
as displayed in Fig.~\ref{fig:Fits}. The fitting parameters are $\delta_a$ and $\delta_f$, which determine the dependence 
of the gap on the magnetic field. The quality of the fits is primarily governed by $\delta_a$ with little sensitivity to $\delta_f$, 
since $\delta_f$ determines the FM order, which mainly arises from Zeeman coupling that is much weaker than on-site repulsion \cite{fitcomment}. 
As $\delta_a$ decreases, corresponding to an increasing strength of subcritical on-site repulsion, the scaling of $\Delta_0$ with $B$ shows a smooth crossover from linear to almost $\sqrt{B}$ scaling. The gap is primarily determined by the easy plane AFM order ($N_\perp$), with very weak accompanying FM order ($m \ll N_\perp$), and we obtain good agreement with experiments even for $m=\lambda=0$ and $\Delta_0=|\vec{N}|$ \cite{supplementary}. Nevertheless, the existence of weak FM order at $\nu=0$ leads to nontrivial consequences for the $\nu=1$ Hall state, which we discuss below. 

The $\delta_a$ values we determine in our fits are consistent with the size of the gap depending on interaction strength (subcritical), expected
to be larger in suspended graphene \cite{yacoby-PRB} than in graphene on a substrate (BN in Refs.~\cite{novoselov-PNAS, kim-recentscaling}).  We also unify the observed linear and sublinear scaling of the gap with magnetic field, illustrating that the scaling form  depends on the strength of short-range interactions.  Our proposal of CSB order in the $\nu = 0$ Hall state can be tested by varying the strength of 
interactions via the following: $(i)$ changing substrates -- higher dielectric constants imply weaker interactions and a more linear field dependence of the gap; $(ii)$ gating MLG (e.g. through a second closely spaced graphene layer with tunable density to screen the primary layer), to vary the strength of interactions {\it in situ}. Figure~\ref{fig:Fits} (inset) shows the dependence of the gap on $\delta_a$ at fixed magnetic field, illustrating this point.
\begin{figure}[htb]
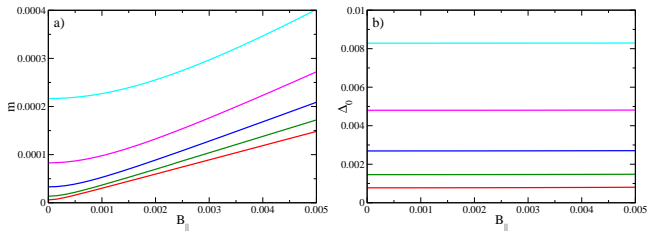

\includegraphics*[width=4.27cm]{Figure2a.eps}
\includegraphics*[width=4.13cm]{Figure2b.eps}
\caption{(Color online) Scaling of (a) FM order ($m$) and (b) total gap ($\Delta_0$) at $\nu=0$ as a function of $B_\parallel$, for $B_\perp = 0.0002$, $0.0004$, $0.0008$, $0.0016$ and $0.0032$ from bottom to top, with $\delta_a = 0.225$ and $\delta_f = 0.05$. $B_\perp$, $B_\parallel$ are measured in units of $B_0$, and $m, \Delta_0$ in units of $E_c$ (see the caption of Fig. 1).}
\label{Fig2}
\end{figure}

The evolution of the $\nu=0$ Hall state in a tilted magnetic field provides an extra constraint on our theory. The perpendicular component of the field ($B_\perp$) gives rise to the Dirac LLs, while the Zeeman coupling scales linearly with the total magnetic field, $B_T$ \cite{klimenko-tilted}. In Fig.~\ref{Fig2} we show the dependence of the FM order, $m$, and the gap $\Delta_0$ on parallel magnetic field for $0$ T $<B_\parallel<50$ T at values of $\delta_a$, $\delta_f$ obtained from our fits to data from Ref.~\cite{kim-recentscaling} at various fixed values of $B_\perp$. $m$ increases roughly \emph{linearly} with $B_\parallel$ for $B_\parallel \gtrsim 2$ T, whereas $N_\perp$ and $\Delta_0$ are insensitive to $B_\parallel$ \cite{supplementary}. Thus we can rule out a \emph{transition} from the easy-plane AFM to a pure FM state with tilting of the magnetic field.

A pure FM state in the ZLL of graphene supports two gapless counter-propagating edge modes, giving a Hall conductivity $\sigma_{xy}=2 e^2/h$ \cite{abanin-edge, miransky-edge}, whereas the easy-plane and pure AFM phases have fully gapped edge modes. However, the gap for the edge modes will 
decrease with increasing FM component in the easy-plane AFM state. Therefore, as $B_{\parallel}$ is increased at fixed $B_\perp$ 
we can expect $R_{xx}$ to decrease, as observed in experiment \cite{kim-recentscaling}.  A measurement of the local density of states
 in the bulk, e.g. with scanning tunneling microscopy (STM), could test our prediction that the total gap $\Delta_0$ is 
insensitive to a tilted magnetic field, providing a strong test of our proposed scenario for the $\nu=0$ Hall state.

In the presence of easy-plane AFM order, the two spin projections in the ZLL are localized on opposite sublattices. Thus, when the chemical potential is placed close to the energy of the first excited state, $\Delta_0$, AFM ordering parallel to the applied magnetic field, 
$N_3$, simultaneously lifts the residual sublattice and staggered spin degeneracy, and gives 
rise to a $\nu=\pm 1$ quantum Hall state. We find the excitation gap for the $\nu=1$ Hall 
state by expanding $E_{0,\sigma}$ in Eq.~(\ref{DiracLL}) to leading order in $N_3$ \cite{herbut-so3}:
$\Delta^{\rm sp}_{1}=2 (m+\lambda) N_3/\Delta_0 + {\cal O} (N^2_3).$
The lifting of sublattice degeneracy by $N_3$ at $\nu=1$ supports CDW ordering ($C$), another CSB order and a natural ground state in graphene in a magnetic field
 for weak nearest-neighbor (NN) repulsion ($V_1$). Therefore, the \emph{total} activation gap for the $\nu=1$ Hall state, $\Delta_1$, will have a spin contribution $\Delta^{\rm sp}_{1}$ from easy-axis AFM ordering and a charge contribution from CDW ordering, i.e. $\Delta_{1}=\Delta^{\rm sp}_{1}+ \Delta^{\rm ch}_{1}$, with $\Delta^{\rm ch}_{1} \equiv C$. Note that $N_3, N_\perp \sim U$ (onsite repulsion), but only half (one-quarter) of the ZLL contributes to the condensation energy in the presence of $N_\perp(N_3)$ order, thus $(N_3/N_\perp) <1$. However, $C \sim V_1 \approx U/2$ in graphene \cite{katsnelson}, and for a subcritical Hubbard interaction $(m+\lambda) \ll N_\perp$, thus $C \gg (m+\lambda)$. Hence, $\Delta^{\rm ch}_{1} \gg \Delta^{\rm sp}_{1}$ and the activation gap for the $\nu=1$ Hall state in a perpendicular magnetic field is dominated by the CDW order, i.e. $\Delta_1 \approx \Delta^{\rm ch}_1 = C$.

With underlying CDW order, the Dirac LLs are at $E^C_n=\pm \sqrt{2 n B +C^2}$ for $n=0,1,2, \cdots$, and for the chemical potential close to the first excited state the free energy is $F_1=\frac{C^2}{4 g_c}-  D \left[ \frac{C}{2} + 2\sum_{n \geq 1} E^C_n\right]$, with $g_c \sim V_1$. Minimizing $F_1$ with respect to $C$, we obtain a self-consistent gap equation exhibiting an ultraviolet divergence because CDW order pushes down all the filled LLs with $n \geq 1$. We regularize this divergence by defining  $\delta_c=\sqrt{\pi} \left[ (g_c \Lambda)^{-1} - (g^c_c \Lambda)^{-1} \right]$, where $(g^c_c)^{-1}=\int^\infty_{\Lambda^{-1}}ds \, s^{-3/2}$ is the critical strength of NN repulsion for CDW ordering at zero magnetic field \cite{roy-scaling}. 
The gap equation is
\begin{equation}
B \int^\infty_0 \frac{ds}{s^{3/2}} \left[ 1- \frac{s e^{-s C^2}}{\tanh{(s B)}} + \frac{s e^{-s C^2}}{2}\right]+\delta_c=0.
\label{eq:nu1gap}
\end{equation}
\begin{figure}[htb]
\includegraphics*[width=8.5cm]{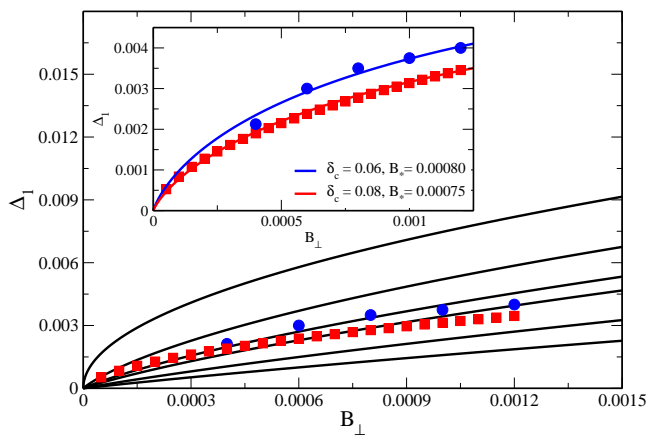}
\caption{(Color online) Scaling of $\Delta_1 = C$ (in units of $E_c$) with perpendicular magnetic field $B$ (in units of $B_0$), for 
$\delta_c=0, 0.03,$ $0.06, 0.08, 0.15,$ and $0.25$ from top to bottom. Red and green dots are the gaps at $\nu=1$ reported in Refs. \onlinecite{yacoby-PRB} and \onlinecite{novoselov-PNAS} respectively. Inset: Scaling of $C$ after incorporating the logarithmic correction to scaling.}
\label{Fig3}
\end{figure}
The scaling of the gap with perpendicular magnetic field  is shown in Fig.~\ref{Fig3} for several values of subcritical NN interactions $(\delta_c>0)$. The calculated gaps are similar in magnitude to those obtained experimentally in Refs.~\cite{novoselov-PNAS} and \cite{yacoby-PRB} for $\delta_c = 0.06$ and $0.08$ respectively. The range of the Coulomb interaction for the CDW ordering at $\nu=1$ is longer than for the easy-plane AFM at $\nu=0$, and hence we take into account the effect of its long range tail, which provides a 
\emph{logarithmic} correction to the Fermi velocity: $v_F=v^0_F \left[1+ \frac{e^2}{ 8 \epsilon v^0_F} \log\left( B_*/B \right) \right]$, with 
$\epsilon$ the dielectric constant of the substrate, $v^0_F$  the bare Fermi velocity, and $1/\sqrt{B_*}$ a characteristic 
length scale for the measured value of $v_F$ \cite{roy-scaling}. Hence, $\Delta_1$ measured in units of $E_c = v_F \Lambda$ acquires a logarithmic correction. Including this correction we obtain excellent agreement with experimentally observed scalings reported in Ref.~\cite{novoselov-PNAS} and Ref.~\cite{yacoby-PRB} for $B_*=8$ T and $7.5$ T, respectively, as shown in Fig.~\ref{Fig3} (inset).

As seen above, CDW order scales with $B_\perp$ but not $B_\parallel$ and hence $\Delta^{\rm ch}_{1}$ is independent of $B_\parallel$. From Fig.~2 it is clear that $m \sim B_\parallel$ while $\Delta_0$ is essentially independent of $B_\parallel$ for fixed $B_\perp$, and $\lambda\sim B_T$, $N_3 \sim B_\perp$. Thus we expect the total activation gap for the $\nu=1$ Hall state, $\Delta_{1}$, to vary \emph{linearly} with $B_\parallel$ for fixed $B_\perp$ since $\Delta^{\rm sp}_1 \sim (m+\lambda)$. Recently, scaling of the gap at $\nu=1$ has been explored in tilted magnetic fields for various fixed values of $B_\perp$, showing \emph{linear} scaling with $B_T$. However, the observed slope is \emph{larger} than that for single-particle Zeeman coupling \cite{kim-recentscaling}. Our proposed scenario for the existence of an underlying spin gap ($\Delta^{sp}_1$) due to finite $N_3$, immediately provides an explanation for the observed \emph{enhanced} slope for the $\nu=1$ Hall state in a tilted magnetic field. 

Alternatively the $\nu=1$ Hall state can form through the generation of two components of \emph{Kekule bond density wave} (KBDW) order, which lift the degeneracy of the ZLL by forming linear combinations of the ZLL states localized on opposite sublattices. These orders, together with the CDW order constitute an $SU(2)$ vector. Hence, a \emph{skyrmionic} excitation of such an $SU(2)$ order can, in principle, be realized at $\nu=1$ \cite{yacoby-PRB}. However, KBDW order can only be realized when NN and next-NN repulsions are comparable \cite{weeks-franz, vozmediano}, while CDW is the natural ground state for NN Coulomb repulsion. In addition, the onsite repulsion supports a finite $N_3$ at $\nu=1$, which in turn gives rise to CDW order. Hence, short-range components of the Coulomb interaction introduce a \emph{strong anisotropy} among the components of this $SU(2)$ vector, which may preempt the appearance of skyrmionic defects at $\nu=1$. The CDW order we proprose at $\nu=1$ could, for example, be established by using sublattice resolved STM measurements, and the easy-axis AFM order could be probed using spin-resolved STM.

To summarize, we address the nature of the broken symmetry phases at $\nu=0, \pm 1$ in MLG, and establish excellent agreement with recent experiments in perpendicular and tilted magnetic fields \cite{yacoby-PRB, kim-recentscaling, novoselov-PNAS}. At $\nu=0$, onsite repulsion leads to easy-plane AFM order and  FM order aligned with the applied field. The $\nu = 0$ state has some superficial similarity with the interlayer canted antiferromagnetic Hall state in bilayer GaAs \cite{sankar-CAF} at even integer filing, which has a different origin from the one in MLG. The ground state at $\nu=1$ simultaneously supports an easy-axis AFM and a CDW, favored by onsite and NN repulsions respectively. Since easy-plane and easy-axis N\'{e}el orders break time-reversal-symmetry, $\nu=0, \pm 1$ Hall states remain robust in the presence of generic time-reversal-symmetric disorders \cite{imry-ma}. We provide several suggestions for experiments to test the scenario we propose. With spin and valley degeneracies completely lifted by the CSB orders the standard Jain sequence of fractional Hall states can be observed for $0 <|\nu|<1$ and $1<|\nu|<2 $ \cite{EAndrei-FQHE, FQHE-Kim-NatPhys,khveshchenko, FQHE-Yacoby-PRL, FQHE-Yacoby-Science}. However, even numerator fractions are less stable than the odd ones when $1<|\nu|<2 $, which may arise from the simultaneous spin and charge ordering in $\nu = 1$ but with a much larger charge gap than spin gap.

B. R. and S. D. S were supported by NSF-JQI-PFC and LPS-CMTC. M. P. K. was supported by NSERC. We thank I. F. Herbut and A. Yacoby for useful discussions and K. S. Novoselov for providing data from Ref. \onlinecite{novoselov-PNAS}.

\onecolumngrid

\vspace{15cm}

\begin{center}
{\bf \large Supplementary Materials for ``Chiral Symmetry Breaking and the Quantum Hall Effect in Monolayer Graphene''} \\
\vspace{0.4cm}
{Bitan Roy$^1$, Malcolm P. Kennett$^2$, and S. Das Sarma$^1$} \\
\vspace{0.25cm}
\emph{$^1$Condensed Matter Theory Center and Joint Quantum Institute, University of Maryland, College Park, Maryland 20742-4111, USA} \\
\emph{$^2$Department of Physics, Simon Fraser University, Burnaby, British Columbia, Canada V5A 1S6}
\end{center}
\vspace{1cm}

In these ``Supplementary Materials", we present additional details, specifically relating to the Hamiltonian 
in the presence of interactions, the calculation of the Landau level spectrum in the presence of chiral 
symmetry breaking orderings, and the derivation of the gap equations discussed in the main text.  We 
also provide some additional numerical results. 

\section{The Landau level spectrum, interaction Hamiltonian and broken symmetry phases in monolayer graphene}

The low energy degrees of freedom of monolayer graphene reside in the two valleys centered on the two 
inequivalent Dirac points  $\pm \vec{K}$ at the edges of the Brillouin zone.  The two valley degrees of
freedom, along with two sublattice and two spin degrees of freedom mean that the states can be 
represented as an 8-component spinor, defined as $\Psi=\left[\Psi_\uparrow, \Psi_\downarrow \right]^\top$, 
where $\Psi^{\top}_\sigma=\big[ u_\sigma (\vec{K}+\vec{q}), 
v_\sigma (\vec{K}+\vec{q}),$  $u_\sigma (-\vec{K}+\vec{q}), 
v_\sigma (-\vec{K}+\vec{q}) \big]$,
with $|\vec{q}| \ll |\vec{K}|$, where  $u_\sigma$ and $v_\sigma$ are the fermionic annihilation operators 
on the two sublattices of the honeycomb lattice and $\sigma=\uparrow, \downarrow$ the two projections of electron spin. 
This spinor representation is invariant under the rotation of electron spin. 
In this basis the Dirac Hamiltonian in the presence of a quantizing magnetic field oriented perpendicular
to the two-dimensional graphene sheet takes the form  
\begin{equation}
H_D[\vec{A}]=\sigma_0 \otimes i \gamma_0 \gamma_j \left(\hat{q}_j - e A_j \right)+ 
\lambda \left( \sigma_3 \otimes I_4 \right),
\end{equation} 
where $\lambda = g\mu_B B$ is the Zeeman coupling, and the magnetic field $\vec{B} =\vec{\nabla} \times \vec{A}$.
The first 2$\times$2 matrix operates on the spin index, whereas the second 4$\times$4 matrix 
acts on valley and sublattice degrees of freedom. 
We choose to work with 4-component $\gamma$ matrices defined as 
$\gamma_0=\sigma_0 \otimes \sigma_3$, $\gamma_1=\sigma_0 \otimes \sigma_2$, 
$\gamma_2=\sigma_3 \otimes \sigma_1$. 
In the absence of Zeeman coupling, the spectrum of $H[\vec{A}]$ is composed of a set of 
Landau levels (LLs) at well separated energies 
$\pm \sqrt{2 n B}$. The degeneracy of the LLs is $1/(\pi l^2_B)$ for $n \geq 1$ and $1/(2 \pi l^2_B)$ for $n=0$.

The short range part of the Coulomb interaction can be represented by a repulsive onsite Hubbard term
\begin{equation}
H_U=\frac{U}{2} \left[ n_\uparrow (A) n_\downarrow (A)+ n_\uparrow (B) n_\downarrow (B) \right],
\end{equation} 
where $n(A/B)_{\sigma}$ represents the fermionic density on sublattice $A/B$ with spin 
projection $\sigma$. The Hubbard interaction can also be written as \cite{herbut-easyplane} 
\begin{equation}
H_U=\frac{U}{16} \bigg[ \left( n(A)+n(B) \right)^2 + \left( n(A)-n(B) \right)^2 - \left( \vec{m}(A)+ \vec{m}(B) \right)^2 -\left( \vec{m}(A)- \vec{m}(B) \right)^2 \bigg],
\label{eq:HubbardU}
\end{equation}
where $n(A)=u^\dagger_\sigma u_\sigma$, and $\vec{m}(A)=u^\dagger_\sigma \vec{\sigma}_{\sigma \sigma'} u_{\sigma'}$ 
are the electronic density and magnetization on the sites of the $A$ sublattice respectively. 
Analogous quantities on the $B$ sublattice are defined in terms of the fermionic operators $v^\dagger_\sigma$ and $v_\sigma$. 
The first term in Eq.~(\ref{eq:HubbardU}) is  the total fermionic density and the second one represents the staggered density. 
The third term gives the total magnetization and the fourth one corresponds to the staggered magnetization. 
Therefore, onsite repulsion can lead to both ferromagnetic (FM) and antiferromagnetic (AFM) orders. 
In the absence of a magnetic field, the appearance of an AFM state at strong Hubbard repulsion 
preempts the formation of a FM state. \cite{herbut-PRL}   However, in the presence of a magnetic field, the Zeeman coupling term 
immediately gives rise to FM order.  We define the order parameters for AFM and FM order as
$\vec{N}=\langle \left( \vec{m}(A)- \vec{m}(B) \right) \rangle$, and $\vec{m}=\langle \left( \vec{m}(A)+ \vec{m}(B) \right) \rangle$ respectively
and set the magnetization to be aligned parallel to the applied magnetic field, i.e. $\vec{m} \to m_3=m$. 
With these two order parameters the effective single particle Dirac Hamiltonian in the presence of the magnetic field is
\begin{equation}
H_{D}= \sigma_0 \otimes i \gamma_0 \gamma_j \left(\hat{q}_j - e A_j \right)+ \left( \lambda +m \right) 
\left( \sigma_3 \otimes I_4 \right) + \left( \vec{N} \cdot \vec{\sigma} \right) \otimes \gamma_0.
\end{equation}  
Diagonalizing this Hamiltonian one obtains the spectrum of the Dirac LLs in the presence of AFM and FM orders, shown in Eq. (1) in the main text. 

\section{Gap equations for the easy-plane anti-ferromagnet phase} 

In this section we present some details of the derivation of the gap equations, shown in Eqs. (5) and (6) of the main text. 
The free energy in the presence of both easy-plane AFM ($N_3=0$) order and FM order is 
\begin{equation}
F_0=\frac{N^2_\perp}{4 g_a}+\frac{m^2}{4 g_f}-D \sum_{\sigma=\pm} \left\{ \frac{1}{2}E_{0,\sigma} + \sum_{n \geq 1} E_{n,\sigma} \right\},
\end{equation}  
as shown in Eq. (2) of the main text, and $E_{n,\sigma} =\sqrt{N^2_\perp+ ([2 n B]^{1/2} + \sigma (m+\lambda))^2}$. 
Minimizing the free energy $F_0$ with respect to $N_\perp$ and $m$ we obtain two coupled gap equations shown in Eq. (3) and (4) of the main part of the paper \begin{eqnarray}
\frac{1}{g_a}  & = &  \frac{B}{\pi} \sum_{\sigma=\pm} \left\{\frac{1}{2E_{0,\sigma}} +  \sum_{n \geq 1} \frac{1}{E_{n,\sigma} } \right\}, \label{eq:gapeq1} \\
\frac{1}{g_f}  & = &  \frac{B}{\pi} \sum_{\sigma=\pm}\left\{ \frac{(m+\lambda)}{2E_{0,\sigma}} + \sum_{n \geq 1} 
\frac{(m+\lambda) + \sigma \sqrt{2 n B}}{E_{n,\sigma}} \right\}. \label{eq:gapeq2}
\end{eqnarray}
As mentioned in the main text, Eq.~(\ref{eq:gapeq1}) has an ultraviolet divergence, for which we now demonstrate the regularization.
Define a function 
\begin{equation}
F(N_\perp, \lambda+m)=\sum_{n \geq 0} \sum_{\sigma=\pm} \frac{1}{\left[ N^2_\perp + \left(\sqrt{2 n B}+\sigma (\lambda+m) \right)^{2} \right]^{1/2}}.
\end{equation}
In terms of this function, Eq.~(\ref{eq:gapeq1}) may be written as 
\begin{equation}
\frac{1}{g_a}= \frac{B}{\pi} \left\{ \left[ F(N_\perp,\lambda+m) -F(N_\perp,0)\right] + F[N_\perp,0] -\frac{1}{\left[ N^2_\perp +(\lambda+m)^2\right]^{1/2}}\right\}.
\end{equation}
The quantity in square brackets is divergence free, and the divergence of the gap equation is captured in $F[N_\perp,0]$. The second
and third terms can be combined to give
\begin{eqnarray}
F[N_\perp,0]-\frac{1}{\left[ N^2_\perp +(\lambda+m)^2\right]^{1/2}}=\frac{1}{\sqrt{\pi}} \int^\infty_{\Lambda^{-2}} \frac{ds}{\sqrt{s}} e^{-s N^2_\perp} \coth(sB)+ \frac{1}{\sqrt{\pi}} \int^\infty_{0} \frac{ds}{\sqrt{s}} e^{-s N^2_\perp} \left[ 1-e^{-s(\lambda+m)^2}\right] ,
\end{eqnarray}
and then Eq. (\ref{eq:gapeq1}) reduces to 
\begin{eqnarray}
\delta_a= \frac{B}{\pi} \bigg[ F(N_\perp,\lambda+m) -F(N_\perp,0) \bigg] -\frac{N_\perp}{\pi^{3/2}} \int^\infty_{0} \frac{ds}{s^{3/2}} \bigg[ 1- s y e^{-s} \coth(s y) \bigg] + \frac{N_\perp}{\pi^{3/2}} \int^\infty_{0} \frac{ds}{\sqrt{s}} e^{-s} \left[ 1-e^{-s x^2}\right],  
\end{eqnarray}
where we have introduced $y=B/N^2_\perp$, $x=(\lambda+m)/N_\perp$, and $\delta_a=1/g_a-1/g^c_a$, with 
$g^c_a$ the critical strength of onsite repulsion for antiferromagnet ordering in zero magnetic field and
\begin{equation}
\frac{1}{g^c_a}=\int^\infty_{\Lambda^{-1}} \frac{ds}{s^{3/2}}.
\end{equation}
On the other hand, the second gap equation, Eq.~(\ref{eq:gapeq2}) does not suffer from any 
ultraviolet divergence. Therefore, the two gap equations, after proper regularization, can be written compactly as 
\begin{eqnarray}
\delta_a- N_\perp y f^a_1(x,y) + \frac{N_\perp}{\sqrt{\pi}} \left(f_2^a(x,y) - yf_3^a(x)\right) & = & 0, \\
\frac{m}{N_\perp} \delta_f-N_\perp y f^m(x,y) & = & 0, 
\end{eqnarray}  
as shown in the main text. The various functions appearing in these two equations are given by 
\begin{eqnarray}
f^a_1(x,y)&=&\sum_{n \geq 0} \sum_{\sigma=\pm} \left[ \frac{1}{\left[ 1+ \left(\sqrt{2 n y} + \sigma x \right)^2 \right]^{1/2}}-\frac{1}{\left( 1+ 2 n y\right)^{1/2}} \right], \nonumber \\
f^a_2(x,y) & = & \int^\infty_0 \frac{ds}{s^{3/2}} \left[1-s y e^{-s} \coth(s y) \right] \nonumber \\
f^a_3(x,y)&=&\int^\infty_0 \frac{ds}{s^{1/2}} e^{-s} \left(1-e^{-s x^2} \right), \nonumber \\
f^m(x,y) & = & \left[\sum_{n \geq 0} \sum_{\sigma=\pm} \frac{\sigma \left(\sqrt{2 n y}+\sigma x \right)}{\left[ 1+ \left(\sqrt{2 n y} +\sigma x \right)^2\right]^{1/2}}\right] 
-\frac{x}{(1+x^2)^{1/2}} .
\end{eqnarray}

\section{Additional numerical results}
\begin{figure}[t]
\includegraphics[height=8cm, angle=-90]{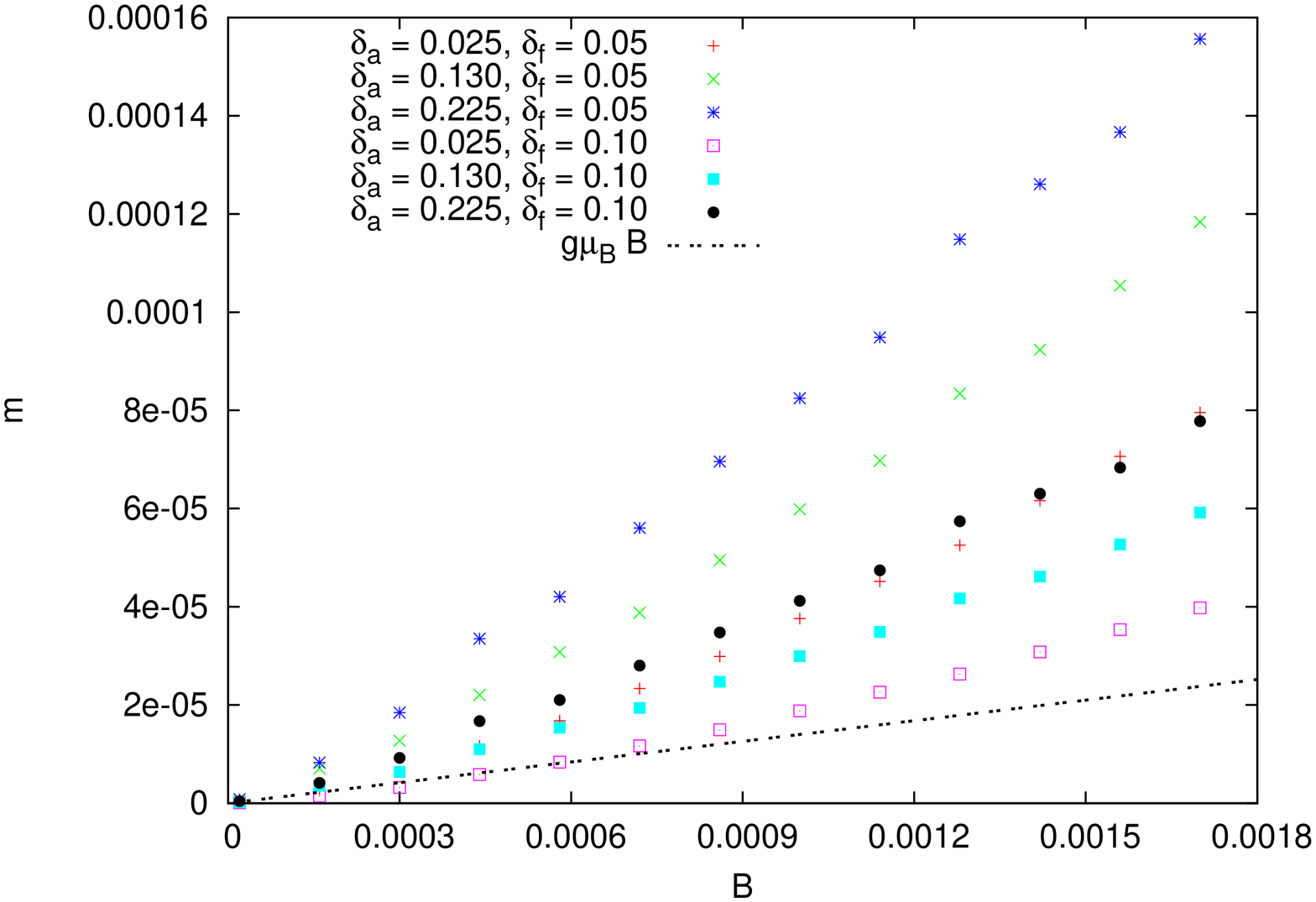} \hspace{0.3cm}
\includegraphics[height=8cm, angle=-90]{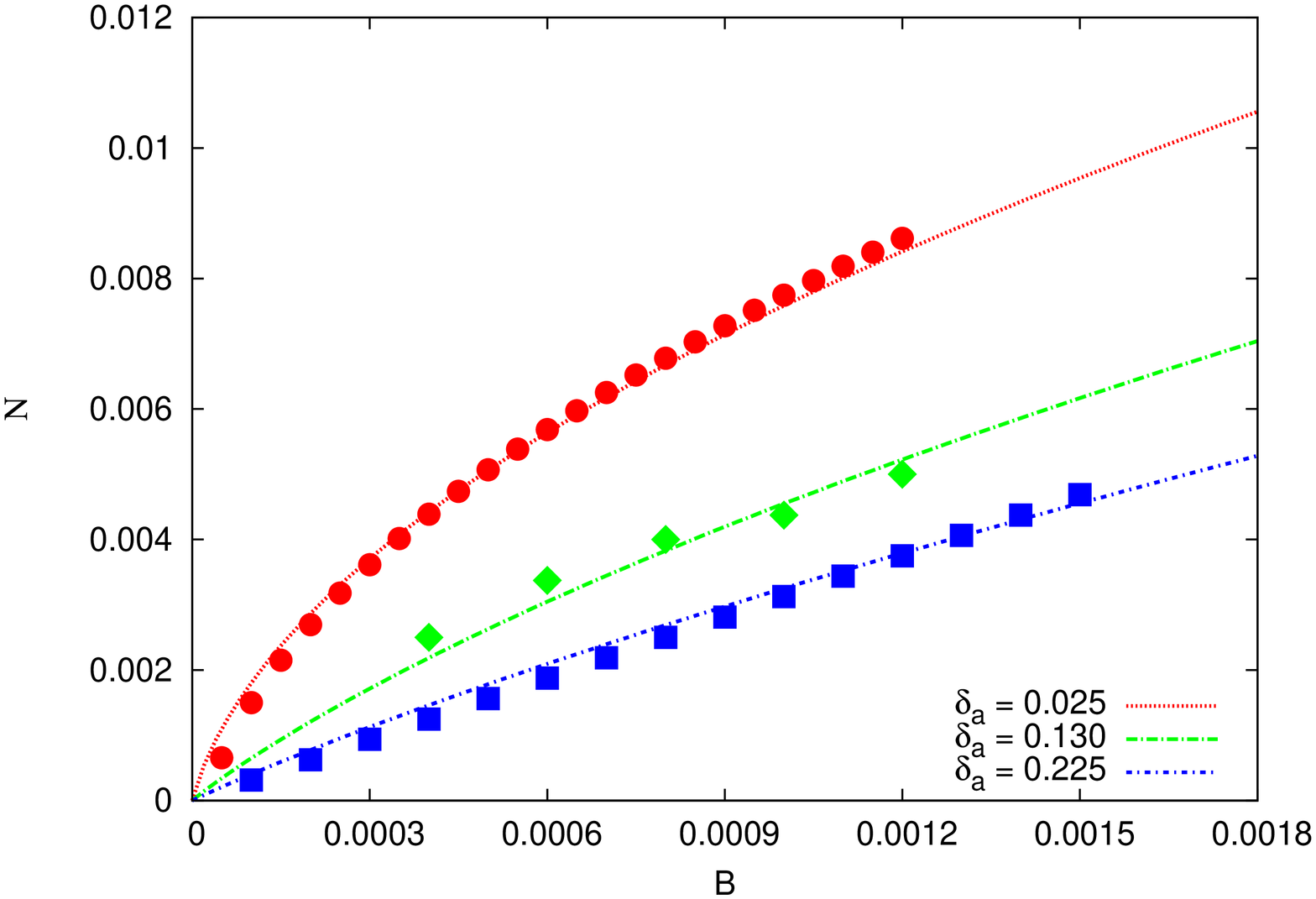}
\caption[] { (Color online) Scaling of FM order ($m$) (left) and easy-plane AFM order ($|N_\perp|=N$) (right) with magnetic field for various choices of the interaction strengths $\delta_a$ and $\delta_f$, quoted in the figures. The dotted line in the left panel 
shows the scaling of the Zeeman coupling with magnetic field. Thus, FM order may be substantially enhanced by onsite Hubbard repulsion, 
but $|N_\perp| \gg (m+\lambda)$. Here $m$ and $N$ are measured in units of $E_c=v_F \Lambda$, and $B$ in units of $B_0\sim \Lambda^2$,
 where $v_F$ is the Fermi velocity and $\Lambda$ is the ultraviolet cutoff for the effective Dirac dispersion in graphene. Red, green, 
blue symbols are experimentally observed gaps reported in Refs.~\onlinecite{yacoby-PRB-supple}, 
\onlinecite{novoselov-PNAS-supple}, and \onlinecite{kim-recentscaling-supple} respectively. Therefore, the right panel shows that the 
fits to experimental data are essentially governed by the easy-plane antiferromagnet order.} 
\label{gaugefieldself}
\end{figure}

We now display some additional numerical results that we have quoted in the main text. We stated that in the easy-plane antiferromagnet state, 
both FM ($m$) and AFM ($|N_\perp|$) orders are non-zero, and for any sub-critical strength of the onsite repulsion, 
$|N_\perp| \gg m$, as illustrated in  Fig.~\ref{gaugefieldself}.

\begin{figure}[htb]
\includegraphics[height=5.0cm, angle=-90]{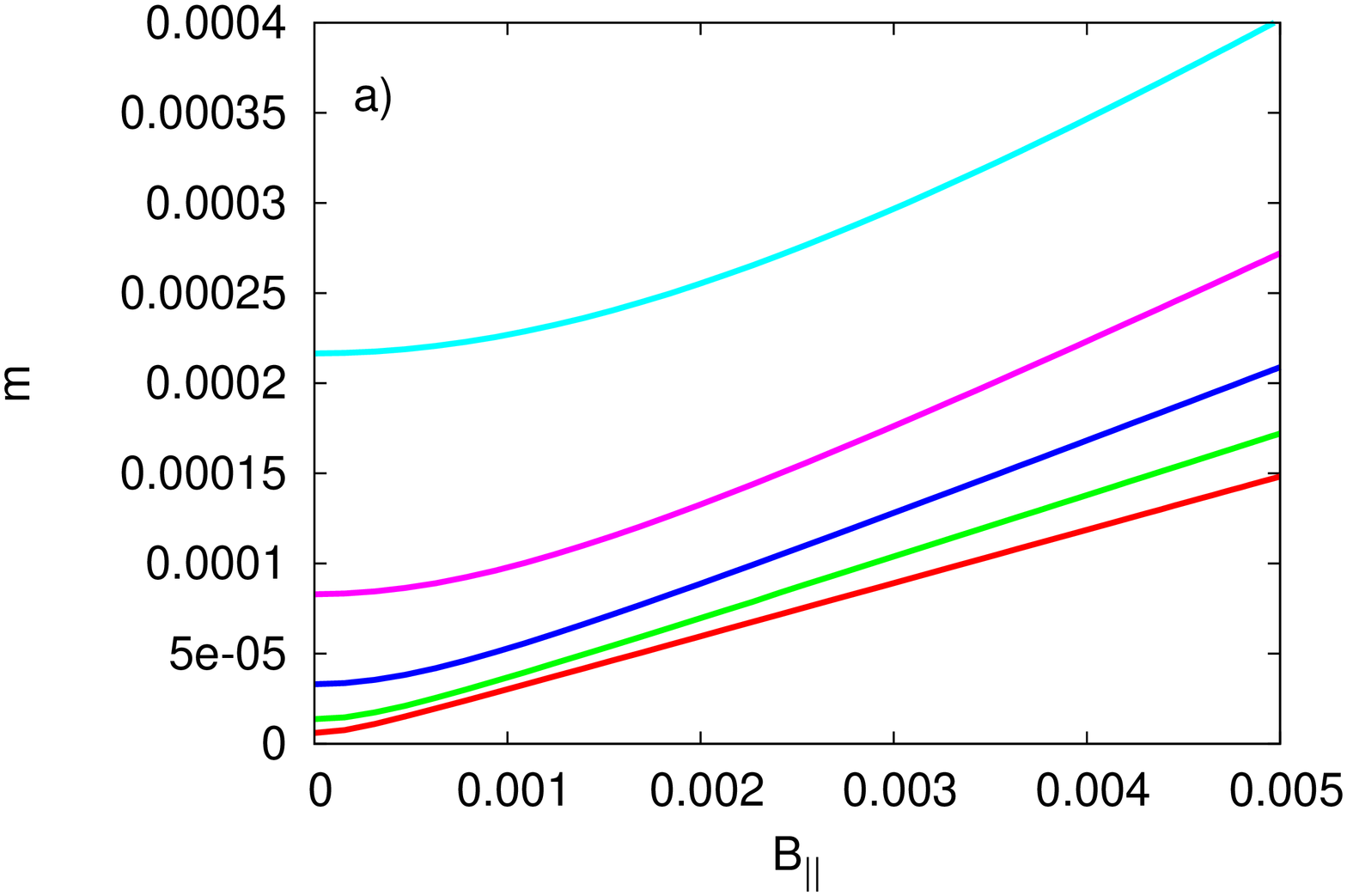}
\includegraphics[height=5.0cm, angle=-90]{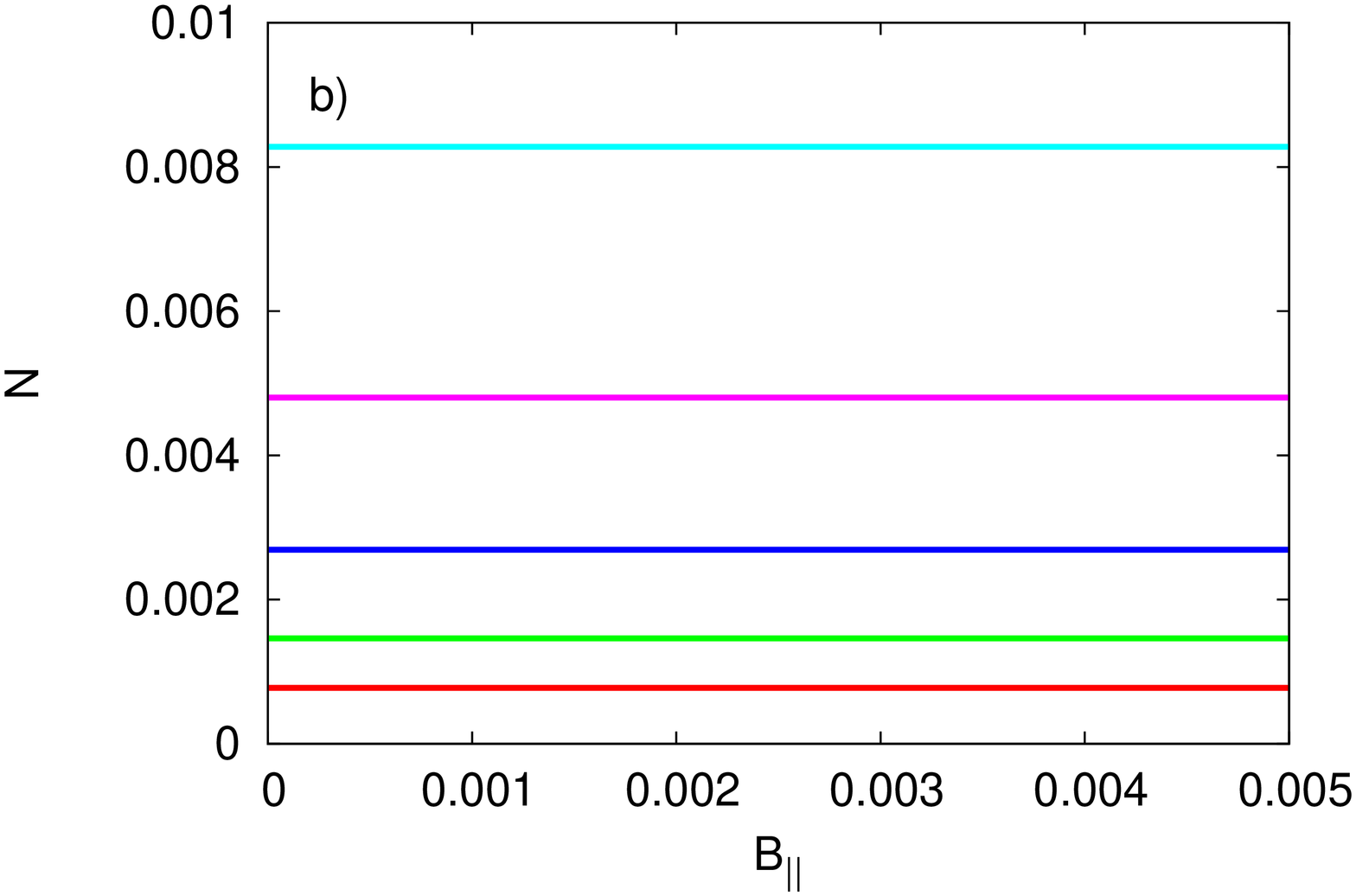}
\includegraphics[height=5.0cm, angle=-90]{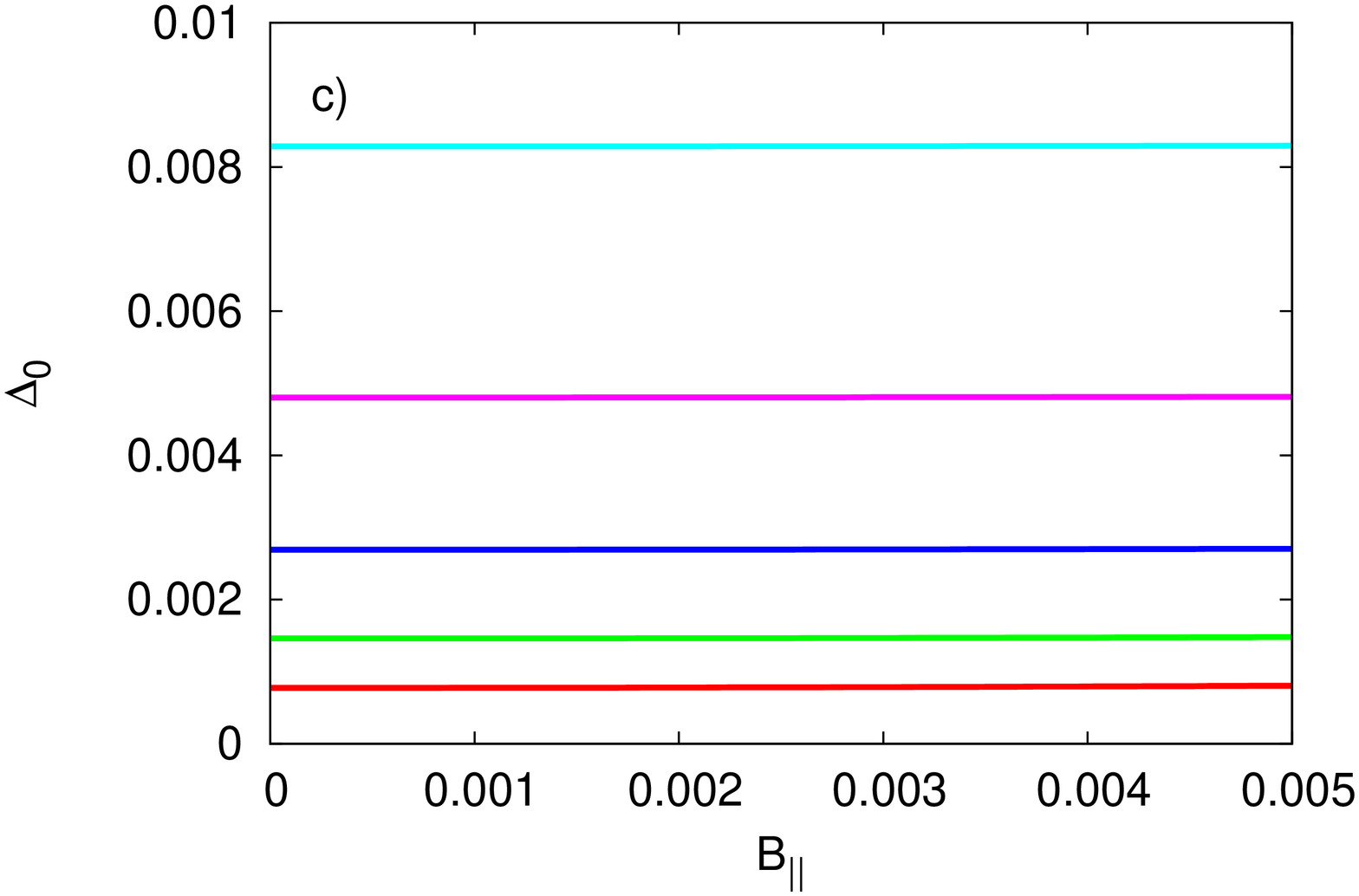}
\caption{(Color online) Scaling of a) FM order, b) AFM order and c) total gap at $\nu=0$ ($\Delta_0$) in a tilted magnetic field, for 
$B_\perp$ = 0.0002, 0.0004, 0.0008, 0.0016 and 0.0032, with $\delta_a = 0.225$ and $\delta_f = 0.05$.  
Here, $B_\perp$ and $B_\parallel$ are measured in units of $B_0$, and $m, \Delta_0$ in units of $E_c$ (see also caption of Fig. 1 of the main text).}
\label{FigS2}
\end{figure}

We now consider the situation when the magnetic field is not perpendicular to the plane of graphene, but oriented in an 
arbitrary direction. In this case, the perpendicular component of the magnetic field gives rise to Dirac LLs, whereas 
the Zeeman splitting couples to the total magnetic field. In the presence of such tilted magnetic fields,
 we showed in the main text that magnetization scales roughly linearly for $B_\parallel >2$ T (recalling that a field of $B = 0.0002 B_0 \sim 2$ T) for 
$B_\perp=2$T, $4$T, $8$T, $16$T and $32$T.   We also showed that the total gap at the charge-neutrality point ($\Delta_0$) 
does not change as one increases $B_\parallel$ for fixed values of $B_\perp$. We also stated that the easy-plane component 
of the AFM order ($N_\perp$) is insensitive to tilting of the magnetic fields. In Fig.~\ref{FigS2} we take the opportunity to display the scaling of these three 
quantities, confirming our statements in the main text.

\begin{thebibliography}{99}
\bibitem{fitcomment} The fit quality decreases noticeably if $\delta_a$ is varied by even $0.01$ from the values shown in Fig.~\ref{fig:Fits}.
\end{thebibliography}
\end{document}